\documentclass[proceedings, preprint]{rmaa}

\usepackage{paralist}
\usepackage{natbib}
\usepackage{psfrag,color}
\usepackage{tablefootnote}
\usepackage{booktabs,caption}
\usepackage[flushleft]{threeparttable}

\SetYear{2024}
\SetConfTitle{The Seventh Workshop on robotic autonomous observatories}

\title{Early-time optical spectral shape measurements of GRB 200925B} 

\author{
  Zhavlonbek Abdullayev,\altaffilmark{1,2} 
  Toktarkhan Komesh,\altaffilmark{1,3,*}
  Bruce Grossan,\altaffilmark{1,4}
  Ernazar Abdikamalov,\altaffilmark{1,5}\\
Zhanat Maksut,\altaffilmark{1}
Maxim Krugov,\altaffilmark{6}
Shynaray Myrzakul\altaffilmark{2}
and Duriya Tuiakbayeva\altaffilmark{1,7}}

\altaffiltext{1}{Energetic Cosmos Laboratory, Nazarbayev University, Astana 010000, Kazakhstan.}

\altaffiltext{2}{Department of General \& Theoretical Physics,
L.N. Gumilyov Eurasian National University, Astana, 010008, Kazakhstan}

\altaffiltext{3}{Institute of Experimental and Theoretical Physics, Al-Farabi Kazakh National University, Almaty, 050040, Kazakhstan}
\altaffiltext{4}{Space Sciences Laboratory, University of California, Berkeley, CA 94720, USA}
\altaffiltext{5}{Department of Physics, School of Sciences and Humanities, Nazarbayev University, 53 Kabanbay Batyr Ave., 010000 Astana}
\altaffiltext{6}{Fesenkov Astrophysical Institute, Almaty, 050020, Kazakhstan}
\altaffiltext{7}{Sh. Ualikhanov Kokshetau University, 76 Abaya str., 020000, Kokshetau, Kazakhstan

*E-mail: toktarkhan.komesh@nu.edu.kz}

\shortauthor{Abdullayev et al.}
\shorttitle{Optical spectral shape measurement of GRB 200925B}

\abstract{
Optical broad-band spectral shape measurements of gamma-ray bursts (GRBs) are typically made starting an hour or more after the trigger event. With our automated, rapid-response system, the Burst Simultaneous Three-channel Imager (BSTI) on the Nazarbayev University Transient Telescope at Assy-Turgen Astrophysical Observatory (NUTTelA-TAO), we began measurements of GRB200925B 129 s after the \textit{Swift} BAT trigger. The temporal decay log slopes in the $g'$, $r'$, and $i'$ bands in the time interval 129 s to 1029 s are -0.43 ± 0.31, -0.43 ± 0.15, and -0.72 ± 0.14, respectively. During the decay phase, a shift in color from red to blue, a change in log slope of  $\beta$ from -2.73 to -1.52 was measured. The evolution in the optical spectral slope is consistent with a decrease in extinction caused by dust destruction.}

\addkeyword{gamma-ray burst: individual: GRB 200925B – dust, extinction.}

\begin{document}
\maketitle

\section{Introduction}
\label{sec:intro}

While the single-band optical flux of gamma-ray bursts (GRBs) has been detected as early as $\sim$10 s after the onset of gamma-ray emission \citep[e.g.,][]{akerlof1999observation, kobayashi2000optical, racusin2008broadband, vestrand2005link, vestrand2014bright, oganesyan2023exceptionally, xin2023prompt, jin2023optical}, the earliest measurements of the spectral shape are typically made much later, starting an hour or more after trigger \citep[e.g.,][]{filgas2011grb,kruhler2011seds,pozanenko2020grb}. In order to study the early-time optical behavior of GRBs, our group developed an automated, rapid-response system, the Burst Simultaneous Three-channel Imager (BSTI) on the 700 mm diameter Nazarbayev University Transient Telescope at Assy-Turgen Astrophysical Observatory \citep{grossan2020performance}. The system responds to \textit{Swift} Burst Alert Telescope (BAT) and other high-quality GRB alert positions and begins simultaneous imaging in $i'$, $r'$, and $g'$ typically 20-100 s after trigger. 

\citet{komesh2023evolution} describes the measurement of the broad-band spectral shape of GRB 201015A with this instrument. The authors found pronounced colour evolution from 58 s to 4140 s after trigger, and concluded that the likely cause was dust destruction due to powerful UV emission in the event. \citet{morgan2014evidence} also found red to blue color evolution in GRB 120119A, but in the initial 200 s following the trigger. 

\begin{figure*}[!t]
  \includegraphics[width=18cm]{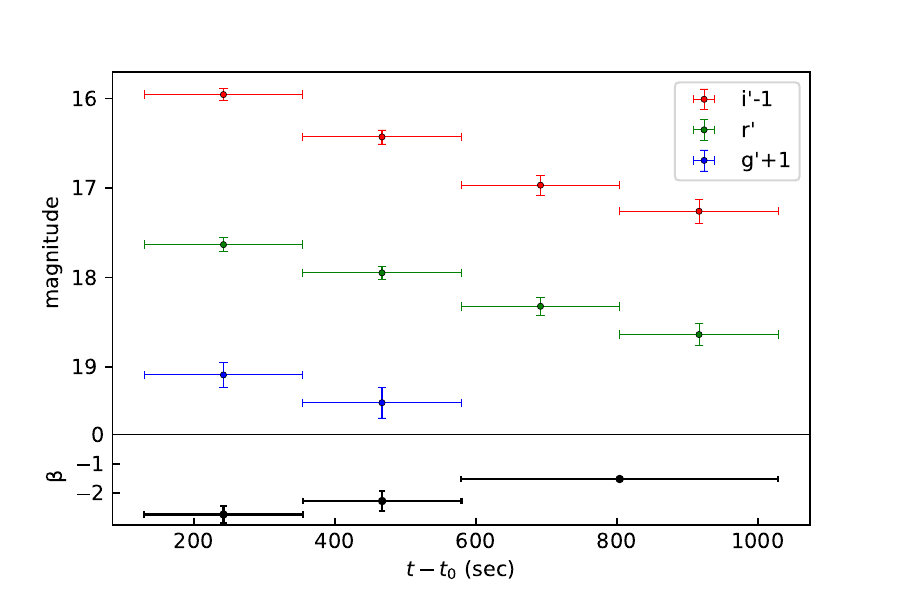}
  \caption{Optical light curve (upper panel) and spectral log slope $\beta$ (lower panel) of GRB 200925B. In this plot, the magnitudes have not been corrected for Galactic extinction. For clarify, we shift the $i'$ and $g'$ bands by -1 and +1 mag, respectively.}
  \label{fig:simple}
\end{figure*}

\begin{table*}
    \caption{NUTTelA-TAO observation of GRB200925B.}
    \label{tab:table3}
    \centering
    \begin{tabular}{cccccccc}
      \toprule
      T$_{mid}$ (s) & t$_{exp}$ (s) & \multicolumn{3}{c}{Magnitudes} & \multicolumn{3}{c}{Flux (mJy)} \\
      \cmidrule(lr){3-5} \cmidrule(lr){6-8}
      & & $g'$ (error) & $r'$ (error) & $i'$ (error) & $g'$ (error) & $r'$ (error) & $i'$ (error) \\
      \midrule
      242 & 225 & 18.08 (0.13) & 17.63 (0.07) & 16.95 (0.06) & 0.23 (0.03) & 0.34 (0.02) & 0.63 (0.04) \\
      467 & 225 & 18.39 (0.17) & 17.94 (0.07) & 17.42 (0.07) & 0.17 (0.02) & 0.26 (0.01) & 0.41 (0.03) \\
      692 & 225 &  UL 18.62             &18.32 (0.01) & 17.96 (0.11) &     UL 0.14         &  0.18 (0.01) & 0.25 (0.02) \\
     1029  & 225 &   UL 18.74            &18.63 (0.12) & 18.25 (0.13) &     UL 0.13          &   0.13 (0.01) & 0.19 (0.02) \\
      \bottomrule
    \end{tabular} 
    \begin{tablenotes}
      \item UL indicates the error value is a 5 sigma upper limit. The mJy flux values are corrected for extinction effects from both within our Milky Way galaxy \citep{schlafly2011measuring} and local to the burst (see Section 3.1 for details).
    \end{tablenotes}
\end{table*}

In this paper, we present and analyze our NUTTelA-TAO optical measurements of GRB 200925B. We include the archival X-$\gamma$ ray measurements in our analysis. Our observations of GRB 200925B began 129 s after the BAT trigger. We observe in the Sloan filter bands $g'$, $r'$, and $i'$ \citep{grossan2020grb}. In the initial phase spanning from 129 s to $\sim$1029 s after trigger, we observe variations in the light curve that differ from those typically observed at later times. These measurements serve as diagnostic tools for probing both the physical mechanisms within the outflow and the environment surrounding the progenitor \citep[e.g.,][]{zaninoni2013gamma,melandri2008early}.

In this paper, we focus on the evolution of the temporal and broad-band spectral slopes during the early afterglow phase. We use the temporal parameter $\alpha$ and the spectral log slope parameter $\beta$ defined via: $f_{\nu} \propto t^{\alpha} \nu^{\beta}$.
Our observations are presented in Section \ref{sec:obs}. Our results are described in Section \ref{sec:results}, and conclusion is provided in Section \ref{sec:conc}. 

\begin{figure*}[!t]
  \includegraphics[width=18cm]{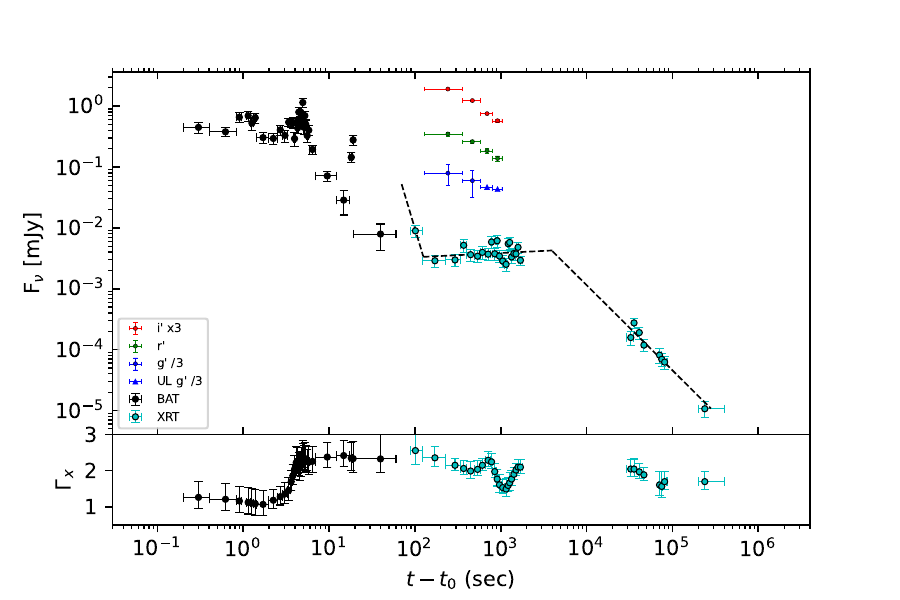}
  \caption{X-ray and optical light curves of GRB200925B. UL refers to the $5 \sigma$ upper limit. BAT refers to the 10 keV  X-ray flux extrapolated from BAT data, and XRT gives the 10 keV flux from XRT data \citep{evans2010swift}. Dashed lines indicate the best fit model ($\chi^2$=29.8, dof=24) with two breaks obtained by SSDCUL \citep{evans2009methods}.} 
  \label{fig:Xray}
\end{figure*}

\begin{figure}[!t]
  \includegraphics[width=0.5\textwidth]{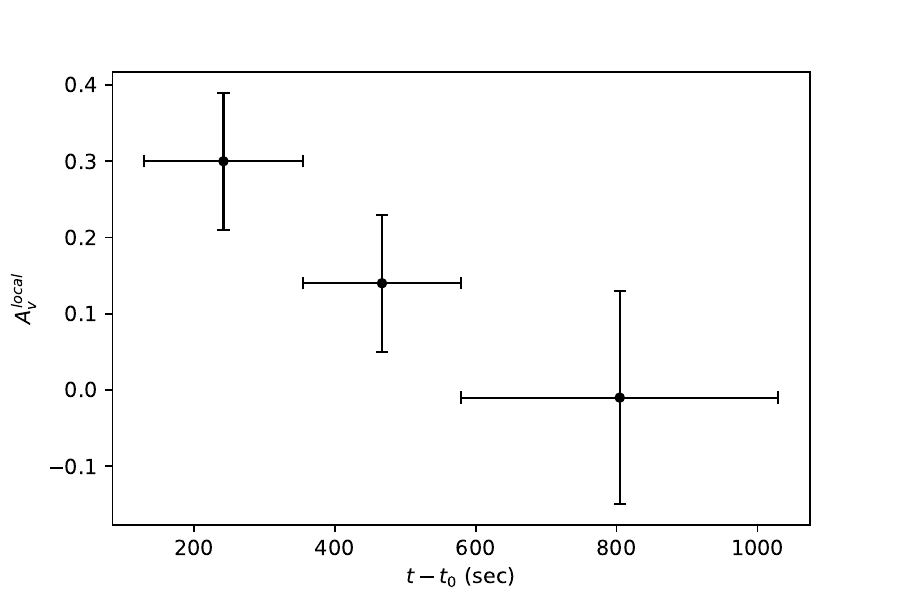}
  \caption{Local extinction, $A_\mathrm{v}^\mathrm{local}$, as a function of time for GRB 200925B. See the main text for definition and method.}
  \label{fig:local_ext}
\end{figure}

\section{Observations}
\label{sec:obs}

We observed GRB 200925B beginning at 21:52:46 UT on September 25, 2020, 129 s after the BAT trigger. Detailed observational parameters can be found in Table \ref{tab:table3}. Our cameras recorded 15 s exposures, but we co-added the images to the exposure time of 225 s in the time interval  129 s to 1029 s after trigger.
The five sigma upper limit sensitivities achieved for the $g'$, $r'$, and $i'$ filter images are 18.5, 19.1 and 18.6 magnitudes, respectively. The weather was clear during the observation. The calibration process was performed using five Pan-STARRS catalog stars identified in our images. We verify the accuracy of our measurements by comparing standard star magnitudes and colors with catalog values. The standard deviation of our magnitude values compared to catalog values of the five standard stars were 0.08, 0.04 and 0.03 mag in the $g'$, $r'$, and $i'$ filters, respectively. We also verify our measurements by examining simulated stars of known actual brightness that we included in our images.

The BAT and XRT data were obtained through the Burst Analyser \citep{evans2010swift}, which is provided by the UK Swift Science Data Centre at the University of Leicester (hereafter "SSDCUL")\footnote{https://www.swift.ac.uk}. 

\section{Results and discussion}
\label{sec:results}

\subsection{Optical light curve and spectral log slope}
\label{sec:EPS}

Figure \ref{fig:simple} (upper panel) shows the optical light curve. In all filters, $g'$, $r'$, and $i'$ , we see a monotonic decay (see Table \ref{tab:table3} for values).  Temporal decay log slopes for the $g'$, $r'$, and $i'$ bands are -0.43 ± 0.31, -0.43 ± 0.15 and -0.72 ± 0.14, respectively. 

To determine the intrinsic log slope of the GRB emission, we account for extinction effects originating from both within our Milky Way galaxy and local surrounding of the burst. The Galactic extinction toward GRB 200925B is 0.133, 0.092 and 0.068 magnitudes in the Sloan $g'$, $r'$ and $i'$ filters, respectively \citep{schlafly2011measuring}.
We use the mean value $A_\mathrm{v}^\mathrm{local}$= 0.15 determined by \citet{li2018large} for the local extinction and the empirical extinction curve from \citet{pei1992interstellar} for the Small Magellanic Cloud (SMC). We obtain the local extinction values for the Sloan $g'$, $r'$ and $i'$ filters as 0.06, 0.04 and 0.03 magnitudes, respectively. 

Figure \ref{fig:simple} (lower panel) shows the spectral log slope as a function of time. The determination of the optical spectral log slope involves fitting a power law spectrum to the $g'$, $r'$ and $i'$ data captured during time interval ranging from 129 s to 580 s and to the $r'$ and $i'$ data captured during time interval of 580 and 1029 s (note that $g'$ is not detected in this time interval).  The obtained spectral log slopes are -2.73 ± 0.28, -2.27 ± 0.32, and -1.52 ± 0.06, respectively.

\subsection{$\gamma$-X-ray light curve}
\label{sec:EPD}

Data of 10 keV values of XRT and BAT were obtained by the SSDCUL. These values were derived through spectral and temporal fits encompassing both the BAT and XRT measurements, ensuring a uniform basis for comparison as described in detail in \citet{evans2007online,evans2009methods,evans2010swift}. 
Figure \ref{fig:Xray} (upper panel) displays the $\gamma$-X-ray light curves alongside the optical light curves. Dashed lines indicate best fit model with 2 breaks \citep{evans2009methods}. Flares and plateau are noticeable in the X-ray light curve during the initial time interval until approximately 1100 s relative to the BAT trigger time.
The early-phase temporal decay log slope is -4.5 (+1.2, -0.9) until 125 s. Subsequently, following the second break at 4100 s, the decay transitions to a log slope of 0.07 (+0.11, -0.12). In the sunsequent phase, it undergoes decay with a temporal decay log slope of -1.4 (+0.25, -0.21).

\subsection{Color Evolution }
\label{sec:CE}

The temporal decay value of longer wavelength band $i'$ is steeper than those in shorter wavelength bands $g'$ and $r'$. This is what one might observe from fading emission behind a dust column that decreased with time, causing decreasing reddening with time. The blue channel would appear to decay slower than the more red ones, which are less affected by reddening. Eventually when the reddening no longer changed, the spectral and temporal slopes would converge to their intrinsic values. Here at late times we do observe these constant slope values, both in the optical and, after 4100 s, in the X-ray. 

In order to determine the changing local reddening, we assume a constant intrinsic spectral and temporal log slope as observed during later stages, with a value of $\beta$=-1.52 and $\alpha$=-0.72. This is the same procedure used in \citet{komesh2023evolution}. Using these intrinsic values, we derive the local reddening as given in 
Fig. \ref{fig:local_ext} shows a decrease in $A_\mathrm{v}^\mathrm{local}$ from 0.30 $\pm$ 0.09 mag to -0.01 $\pm$ 0.14 mag in the time interval 129 s to 580 s.  This amounts to a change of 1.87 $\sigma$. Subsequently, $A_\mathrm{v}^\mathrm{local}$ appears to remain constant.

Long GRBs are observed in regions with active star formation \citep{woosley1993gamma, woosley2006supernova}. The majority of GRBs exhibit a moderate level of dust extinction around $\sim1000$ s, with occasional cases of heavier extinction  \citet{perley2009host}. Red-to-blue color evolution of the color of GRBs has been observed in GRB and attributed to destruction of dust grains in at least the two cases given in the introduction. We observed a monotonic red-to-blue color evolution which is \textit{consistent} with dust destruction, but the weakness of the statistical significance does not allow us to make a definite conclusion that it is present.

\section{Conclusion}
\label{sec:conc}

We studied the early-time optical behavior of GRB 200925B, starting 129 s after the BAT trigger. Our analysis revealed variations in the light curve from 129 s to approximately 1029 s after the trigger. We accounted for extinction effects from both within our Milky Way galaxy and the local environment of the burst. The spectral log slope analysis revealed monotonic change from red to blue during the early afterglow phase. This is \textit{consistent} with decreasing reddening with time, but this change had low statistical significance. Consistency with dust destruction is conceptually significant, however, as we have only observed and analyzed a small number (total 4 including soon-to-be-published) of early light curves, and one was a clear case \citep{komesh2023evolution}, and one consistent with dust destruction. If this trend continues, it will become clear that this is a common and important process. 

Additionally, we presented X-ray and optical light curves, highlighting flares and plateaus in the X-ray light curve during the initial time interval until around 1100 s relative to the BAT trigger time. The temporal decay log slope exhibited changes during different phases of the afterglow.

\section{Ackowledgements}

This research has been funded by the Science Committee of the Ministry of Science and Higher Education of the Republic of Kazakhstan (Grant Nos. AP14870504) and partially supported by the Nazarbayev University Faculty Development Competitive Research Grant Program (No 11022021FD2912). MK is supported by the Science Committee of the Ministry of Science and Higher Education of the Republic of Kazakhstan (Grant No. BR21881880). The NUTTelA-TAO Team expresses gratitude to the personnel of the Assy-Turgen Astrophysical Observatory in Almaty, Kazakhstan, and the Fesenkov Astrophysical Institute, Almaty, Kazakhstan, for their assistance. Special appreciation is extended to the ATO staff. The research utilized data provided by the UK Swift Science Data Centre at the University of Leicester

\bibliography{rm-extenso}

\end{document}